\documentclass[lettersize,journal]{IEEEtran}
\IEEEoverridecommandlockouts

\usepackage{tikz}
\usepackage{url} 
\usepackage{textcomp}
\usepackage{multirow}
\usepackage{tcolorbox}
\usepackage{booktabs}
\usepackage{hyperref}
\usepackage{pifont}
\usepackage{array}

\usepackage{footnote} 

\usepackage[ruled,vlined]{algorithm2e} 
\usepackage{listings} 

\usepackage[table]{xcolor}  
\usepackage{rotating}       

\definecolor{lightgreen}{RGB}{144, 238, 144} 
\usepackage{enumitem}
\usepackage{etoolbox}

\definecolor{MyTitleBackColor}{RGB}{77,77,77}
\definecolor{MyBoxBackColor}{RGB}{238,238,238} 

\newtcolorbox{promptbox}[1]{
    colback=MyBoxBackColor,       
    colframe=MyTitleBackColor,    
    colbacktitle=MyTitleBackColor, 
    coltitle=white,                
    fonttitle=\bfseries,           
    arc=2mm,                       
    boxrule=0.8pt,                 
    boxsep=1.8mm,                    
    top=2mm, bottom=2mm, left=2mm, right=2mm, 
    title=#1,                      
}

\newcommand{\myname}{\textsc{SyLares}\xspace}

\usepackage{cite}
\usepackage{amsmath,amssymb,amsfonts}
\usepackage{graphicx}
\usepackage{textcomp}
\def\BibTeX{{\rm B\kern-.05em{\sc i\kern-.025em b}\kern-.08em
    T\kern-.1667em\lower.7ex\hbox{E}\kern-.125emX}}
\begin{document}

\title{A Neuro-Symbolic Reasoning Engine for Compilation-Free Patch Presence Testing}

\author{Siyuan Li, Jiayi Zhou, Zhiyuan Xia, Weijie Wang, Yaowen Zheng, Weiran Tang, Hong Li, Hongsong Zhu, Shanqing Guo

\thanks{This work was supported by National Key Research and Development Program of China (Grant No.2022YFB3103904)). (Corresponding authors: Hong Li, Siyuan Li)}
\thanks{Siyuan Li, Jiayi Zhou, Zhiyuan Xia and Shanqing Guo is with the School of Cyber Science and Technology, Shandong University, Qingdao 266237, China. (e-mail: siyuan@sdu.edu.cn, 294030000@qq.com, 3325894735@qq.com, guoshanqing@sdu.edu.cn)}
\thanks{Weijie Wang, Yaowen Zheng, Weiran Tang, Hong Li, and Hongsong Zhu are with the Institute of Information Engineering, Chinese Academy of Sciences, Beijing 100085, China, and also with the School of Cyber Security, University of Chinese Academy of Sciences, Beijing 100049, China. (e-mail: \{wangweijie, zhengyaowen, tangweiran, lihong, zhuhongsong\}@iie.ac.cn)}
}

\markboth{Journal of \LaTeX\ Class Files,~Vol.~14, No.~8, August~2021}%
{Shell \MakeLowercase{\textit{et al.}}: A Sample Article Using IEEEtran.cls for IEEE Journals}

\maketitle

\begin{abstract}
Code reuse in modern software ecosystems facilitates the spread of 1-day vulnerabilities, posing significant security risks. Determining if patches are present is crucial, but traditional methods suffer from limited usability and accuracy, often depending on a fragile compilation process.
In our prior work, we introduced Lares, the first compilation-free framework for patch presence testing that combines large language models (LLMs) with SMT solvers. However, Lares treats the LLM and the SMT solver as two \emph{separate, parallel decision channels}: the symbolic channel (Z3) succeeded in only 20\% of code cases, and the remaining cases fell back to the neural channel (the LLM) for the final decision. Letting the LLM act as the final arbiter turns verification into an informal, non-deterministic process that is prone to hallucination errors, while the two channels never reinforce each other.
To overcome these limitations, we propose \myname, which reformulates compilation-free patch presence testing as \emph{neuro-symbolic reasoning}. Instead of letting the LLM emit a final verdict, \myname uses the LLM as a \emph{neural feature extractor} that perceives heterogeneous source code and decompiled pseudocode and produces specification-compliant SMT-LIB inputs, and delegates the final, deterministic judgment to the symbolic Z3 solver. This neural-perceives, symbolic-decides coupling is realized by a Code2SMT reasoning engine that combines deterministic syntactic parsing with LLM-driven code normalization and symbol alignment, so that every case is reduced to a formal equivalence query rather than an opaque LLM guess.
Our evaluation shows that \myname improves accuracy by 10\% and the F1 score by 6\% compared to Lares. Notably, by routing the neural output into the symbolic reasoner, its formal verification coverage reached 100\%, with the Z3 solver processing all test cases, whereas Lares succeeded in only 20\% of cases. Compared to other state-of-the-art methods, \myname achieves a 15\% improvement in F1 score. These findings indicate the significant potential of neuro-symbolic reasoning in the domain of code verification.
\end{abstract}

\begin{IEEEkeywords}
Patch Presence Testing, Binary Analysis, Neuro-Symbolic Reasoning
\end{IEEEkeywords}

\section{Introduction}

\IEEEPARstart{S}{oftware} development is not a purely repetitive activity. To speed up implementation, developers frequently incorporate third-party libraries, creating a complex software supply chain ecosystem \cite{FirmSec}. This practice, however, can propagate vulnerabilities and increase security risk. According to the Synopsys report \cite{Blackduck}, 96\% of evaluated projects used at least one third-party library, 89\% of codebases contained open-source components that had not been updated for over two years, and some for more than four years. Delayed updates and patching amplify the impact of 1-day vulnerabilities.

\begin{figure}

\centering
    \includegraphics[width=\linewidth]{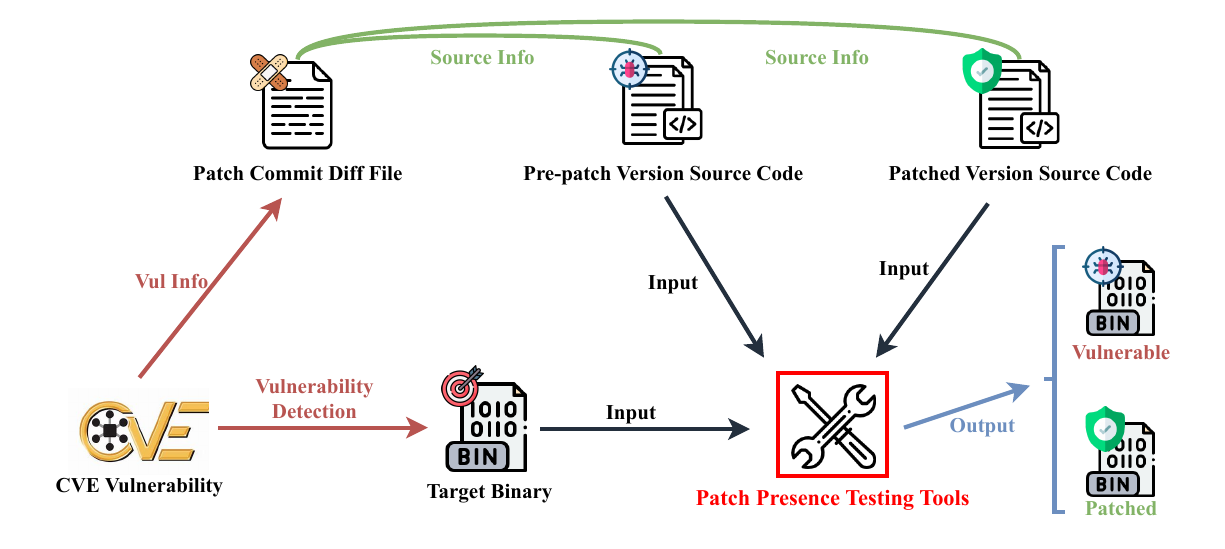}

\caption{The Background.}
\label{fig1}
\end{figure}

Several approaches \cite{Gemini, jTrans, Vulhawk, FirmSec, LibDB, LibAM, binaryai} have been proposed for detecting 1-day vulnerabilities. As shown in Figure~\ref{fig1}, these methods, which fall into syntactic-based and semantic-based categories, typically perform function-level or library-level matching using binary similarity techniques. However, they face significant limitations in real-world scenarios. We summarize these limitations into three main points: First, they rely on a compilation process that is difficult to automate and often fails, reducing usability (\textbf{P1}). Second, they struggle to accurately locate patch-related code fragments, which reduces detection accuracy (\textbf{P2}). Third, their accuracy degrades significantly across different compilation environments, such as varied optimization options or architectures (\textbf{P3}).

In previous work, we introduced Lares \cite{lares}, a novel framework designed to address these challenges. Lares was the first to leverage Large Language Models (LLMs) for patch presence testing, eliminating the need for compilation by extracting features directly from source code and patches (\textbf{for P1}). It employed an LLM to locate patch-related code slices in decompiled pseudocode, ensuring a consistent comparison granularity (\textbf{for P2}), and its features were architecture-independent (\textbf{for P3}). Lares combined formal methods with LLM-driven analysis to detect if a patch was present.

Despite its innovations, Lares is fundamentally limited by \emph{how} it organizes the LLM and the SMT solver: it treats them as two \emph{separate, parallel decision channels} rather than as one coupled reasoning process. The symbolic channel (Z3) and the neural channel (the LLM) each attempt to reach a verdict independently, and the system simply falls back from one to the other. This decoupled design causes two critical challenges. First, the symbolic channel struggles with complex control flows and data structures, exhibiting limited applicability, and succeeds in only 20\% of code cases, significantly reducing judgment's reliability (\textbf{Challenge 1, C1}). Second, whenever the symbolic channel fails, the neural channel must issue the final verdict; this LLM-as-final-arbiter judgment is an informal, non-deterministic verification process that is highly susceptible to hallucination, leading to erroneous conclusions (\textbf{Challenge 2, C2}). Because the two channels never reinforce each other, Lares produced a high rate of false positives, undermining its effectiveness in practice.

To overcome these challenges, we propose \myname, which reformulates compilation-free patch presence testing as \emph{neuro-symbolic reasoning}: instead of letting the LLM and the solver compete as parallel channels, \myname couples them so that the \emph{neural} component perceives and the \emph{symbolic} component decides. The key idea is to use the LLM as a feature extractor rather than the final discriminator, and to delegate the final, deterministic judgment to the symbolic Z3 solver. Concretely, the patch enhancement module first expands patch semantics by adding context-related statements to the patch code with a few statements, and the patch localization module leverages an LLM to identify pseudocode slices corresponding to patch code slices within the target function's pseudocode. These localized and enhanced code slices then serve as the input for the Code2SMT reasoning engine, which realizes the neuro-symbolic coupling in three stages. Its Extractor utilizes an FSM-based parser for rapid lexical analysis, decomposing the code into fundamental logical units, and employs an LLM-based parser as a robust fallback for syntactically complex expressions. Its Normalizer then uses the LLM's code analysis and translation capabilities not to judge, but to translate these heterogeneous units into a specification-compliant SMT-LIB representation and to align symbols across source code and pseudocode. Finally, the symbolic Verifier feeds this normalized representation into the Z3 solver, which makes the final deterministic decision. By turning the LLM's flexible perception into well-formed symbolic inputs, \myname broadens the applicability of the SMT solver (\textbf{for C1}) and eliminates the need for an LLM-based final judgment, thereby mitigating its associated hallucinations (\textbf{for C2}).

We thoroughly analyze the issues introduced by the compilation process in existing methods and demonstrate how \myname naturally avoids them. Our evaluation demonstrates that \myname significantly enhances the accuracy and reliability of patch presence testing. Compared to Lares, our system improves accuracy by 10\% and the F1 score by 6\%. Our experiments show that routing the LLM's neural output into the symbolic reasoner effectively resolves the critical failure point of Lares, where the SMT solver previously succeeded in only 20\% of cases (addressing C1). A key point is the expanded capability of the SMT solver, which achieved 100\% formal verification coverage by processing all test cases. As a direct result, delegating the final decision to the symbolic reasoner significantly reduces the false positives caused by the LLM hallucinations inherent in Lares's LLM-based judgment (addressing C2). Furthermore, when compared against other methods, \myname achieves a 20\% improvement in precision and a 15\% improvement in F1 score across a diverse range of architectures and compilation options.

We summarize our main contributions below:

\begin{figure*}
\centering
    \includegraphics[width=\linewidth]{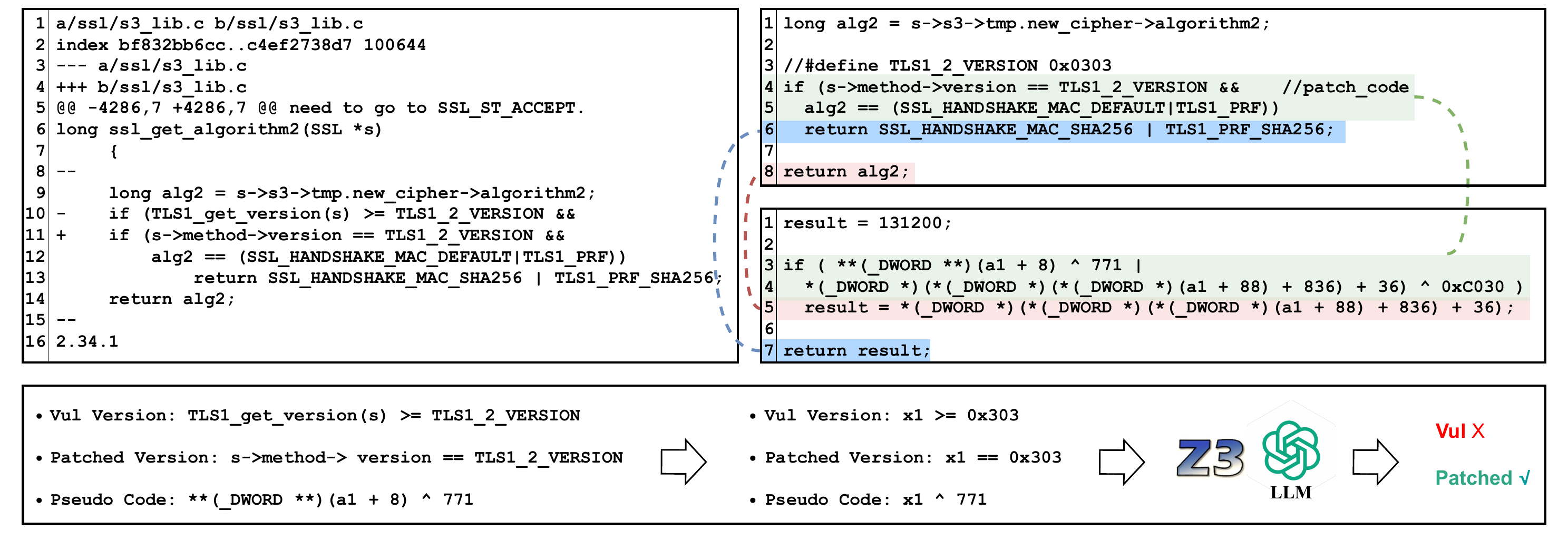}

\caption{A motivation example.}
\label{fig:motivation}

\end{figure*}

\begin{itemize}

\item We propose \myname, the first compilation-free, neuro-symbolic approach for patch presence testing, which eliminates substantial manual work and achieves large-scale automated detection.

\item We identify that coupling, rather than merely combining, the neural and symbolic components is the key to reliable verification, and we design and implement a Code2SMT reasoning engine in which the LLM serves as a neural feature extractor that produces specification-compliant SMT-LIB inputs and the Z3 solver serves as the deterministic decision maker, achieving a 100\% formal verification rate.

\item We implement \myname with 3,500 lines of Python code and integrate it with several source and binary analysis tools, while also validating its performance using advanced LLMs. The code and dataset are available at \url{https://github.com/Siyuan-Li201/SyLares}

\item We conduct an extensive evaluation of \myname, and the results show that \myname outperforms state-of-the-art approaches in terms of both accuracy and recall overall across a range of architectures and compilation options.

\end{itemize}

\section{Motivation}

We illustrate our motivation and key insights using the example in Figure~\ref{fig:motivation}. First, we show why patch presence testing is necessary. Then, we examine the limitations of existing methods and outline insights that guide the design of more robust tools for patch presence testing.

\subsection{Necessity of Patch Presence Testing}
Figure~\ref{fig:motivation} presents CVE-2013-6449 as a representative case. The leftmost box shows the patch commit, which makes small edits: it replaces a version-related macro with a structure pointer and changes the comparison operator from ``\textgreater=" to ``==". 
These minor edits leave the patched function very similar to the vulnerable one. As a result, existing binary similarity methods, which focus on function-level matches, have difficulty distinguishing the two versions. This underscores the need for a finer-grained patch existence verification method to detect the vulnerability accurately.

\subsection{Insight 1: The Need for a Compilation-Free Approach}
Existing approaches depend on compilation to extract features such as control flow graphs (CFG) \cite{BinXray}, assembly code \cite{robin, ps3}, or intermediate representations (IR) \cite{react}. For each vulnerability, they require manual compilation of both pre-patch and patched binaries, along with debugging information.

\begin{table}
  \centering
  \caption{Usability Evaluation of Existing Methods}
  \vspace{0.5em}
  \label{tab:usability}
  \setlength{\tabcolsep}{4pt}
  \begin{tabular}{cccccc}
    \toprule
     Problems & REACT & BinXray & Robin &  PS3 & \myname\\
    \midrule
    ED & \ding{56} &  \ding{56} & \ding{56} & \ding{56} & \ding{52}\\
    CP & \ding{56} &  \ding{56} & \ding{56} & \ding{56} & \ding{52}\\
    CV & \ding{56} & \ding{56} &  \ding{56} & \ding{56} & \ding{52}\\
    BT & \ding{56} & \ding{56} &  \ding{56} & \ding{56} & \ding{52}\\
    AC & \ding{56} & \ding{52} &  \ding{52} & \ding{52} & \ding{52}\\
  \bottomrule
  \vspace{-1.0em}
\end{tabular}

\smallskip
\noindent\small 
\textit{\textbf{Note:}} ED: External Dependencies. CP: Compilation Parameters. CV: Compiler Versions. BT: Build Tools. AC: Additional Compilation. \ding{52} indicates that this method can solve the problem, while \ding{56} indicates that it cannot.

\end{table}

However, as summarized in Table~\ref{tab:usability}, this reliance on compilation is a critical bottleneck. In practice, compilation is difficult to automate and often demands substantial domain knowledge; many projects fail even with manual effort due to complex dependencies, compiler/build tool mismatches, and parameter configuration.

Therefore, Lares proposed a compile-free approach for patch presence testing. Unlike prior methods, Lares decompiles the target binary to pseudocode and analyzes it by extracting features from the source code for comparison. This design inherently avoids all issues tied to the fragile compilation process.

\subsection{Insight 2: From Decoupled Channels to Neuro-Symbolic Reasoning}
\begin{figure*}
\centering
    \makebox[\textwidth][c]{%
        \includegraphics[width=0.9\textwidth]{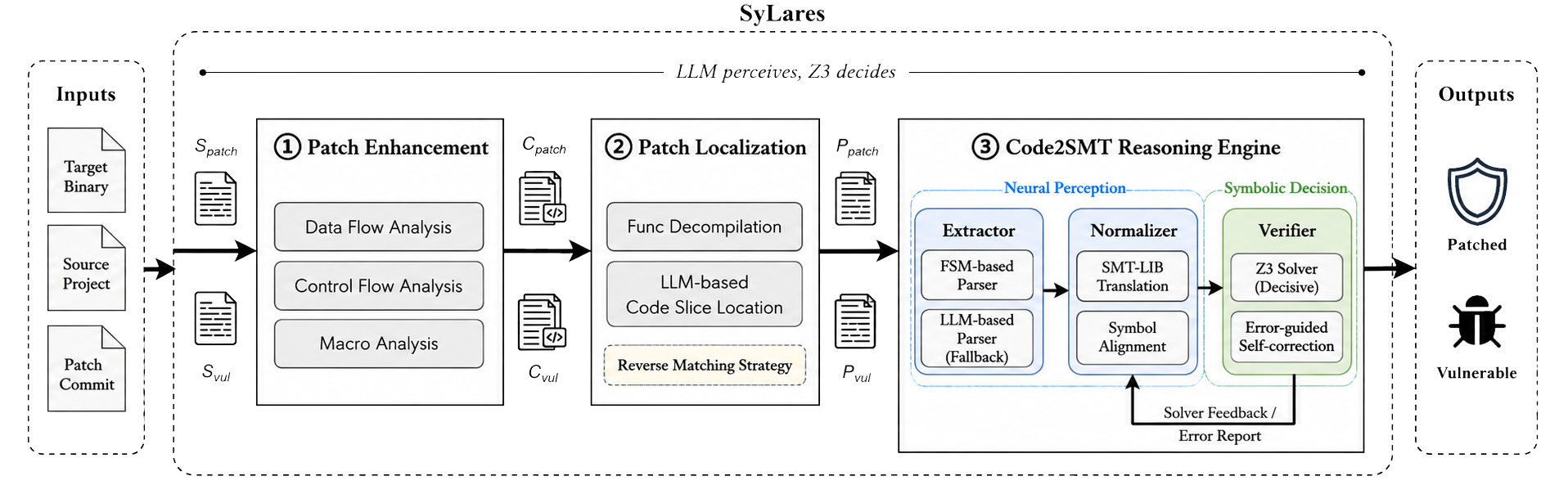} 
    }

\caption{The workflow of SyLares.}
\label{fig:workflow}

\end{figure*}

Even after accurately localizing a code slice, robustly verifying its semantics remains a significant challenge. The ultimate goal is to determine if the semantics of the pseudocode slice in the target binary align with the pre-patch or post-patch source code. Two capabilities are needed at once: the \emph{perceptual} ability to understand heterogeneous, noisy code representations (C source and decompiled pseudocode), and the \emph{deductive} ability to reach a sound, deterministic verdict. Neither an LLM nor an SMT solver alone covers both. While SMT solvers like Z3 provide the required logical precision, they cannot directly compare disparate code representations; they operate only on well-formed, canonical logical formulas. LLMs, conversely, perceive code flexibly but reason non-deterministically and hallucinate.

Lares attempts to cover this gap by running the two as separate, parallel channels: it first tries the symbolic channel (Z3) and, on failure, falls back to the neural channel (the LLM) for the final decision. The hidden cost of this decoupling is a translation gap that neither channel bridges. Decompilers often produce complex and non-standard code structures, including intricate pointer arithmetic, memory layout-dependent variable accesses (e.g., **(\_DWORD **)(a1 + 8)), and convoluted expressions that do not conform to standard C syntax. As illustrated at the bottom of Figure~\ref{fig:motivation}, a simple comparison like s$\rightarrow$method$\rightarrow$version == TLS1\_2\_VERSION becomes **(DWORD **)(a1 + 8)\verb|^|771 in the pseudocode. However, the traditional parser used in Lares cannot handle such convoluted, non-standard expressions, severely limiting the applicability of the symbolic channel.

Our analysis shows that this translation bottleneck rendered the symbolic channel in Lares ineffective, succeeding in only 20\% of complex code cases. This critical failure forces Lares to degenerate into its neural channel for the majority of cases, relying solely on an LLM as the final arbiter of semantic judgment, which we define as an informal verification process. As illustrated in the Lares workflow in Figure \ref{fig:motivation}, the SMT solver fails (reporting Not Sure), forcing the LLM to make the final judgment. This process is highly susceptible to LLM hallucinations: the LLM Judgement Result box highlights this failure. The model, unable to semantically resolve the ambiguous x1\verb|^|771 artifact (Points 5), provides a flawed justification based on nonsensical heuristics, such as `Equality is common and may be unrelated' (Point 6) and that the prefix logic `best fits permissive... behavior' (Point 7). This flawed reasoning leads directly to a false positive conclusion, undermining the reliability of the entire system.

This motivates our core insight: rather than letting the neural and symbolic components act as competing channels, we couple them into a single neuro-symbolic reasoning process in which the LLM \emph{perceives} and the solver \emph{decides}. As shown in the \myname workflow in Figure \ref{fig:motivation}, we propose a Code2SMT reasoning engine. Here the LLM is repositioned as a neural feature extractor rather than the final discriminator: it uses its flexible code understanding not to replace the solver, but to enable it, by robustly translating the non-standard pseudocode (e.g., x1\textasciicircum{}771) into its correct logical equivalent (e.g., x1 == 0x0303) and aligning symbols across the two representations. This specification-compliant SMT-LIB form is then handed to the symbolic Z3 solver, which makes the final deterministic judgment. By avoiding any non-deterministic LLM verdict and channeling neural perception into symbolic deduction, this design significantly expands the reach of formal verification and achieves a reliable patch presence test.

\section{Methodology}


In this section, we introduce the design of \myname. We begin by explaining its overall workflow, followed by a detailed description of each module.

\subsection{Overview}

The \myname framework, as shown in Figure~\ref{fig:workflow}, employs a three-stage pipeline to verify the patch status of a given vulnerability. The pipeline comprises: (1) Patch Enhancement, (2) Patch Localization, and (3) neuro-symbolic reasoning via a Code2SMT reasoning engine. Conceptually, the first two stages constitute the \emph{neural perception} front-end, where the LLM is used to understand and prepare heterogeneous code, while the third stage constitutes the \emph{symbolic decision} back-end, where the Z3 solver renders the final deterministic verdict. The LLM never issues the verdict itself; it only feeds specification-compliant inputs to the solver. Our methodology operates on the premise that a potentially vulnerable function $F_{t}$ has been previously identified within a target binary by a code-similarity detection scheme (e.g., BinaryAI \cite{binaryai}). Taking the target binary, its corresponding source project, and the relevant patch commit as inputs, \myname formally determines whether the vulnerability has been patched.

\textbf{Patch Enhancement}. 
The primary function of this module is to process the source code of a vulnerable function, $S_{vul}$, and its patched counterpart, $S_{patch}$. Its objective is to extract semantically relevant code slices, denoted $C_{vul}$ and $C_{patch}$, to facilitate subsequent patch localization within the target. Security patches often consist of minimal syntactic changes, rendering a simple textual diff insufficient to capture the full semantic context. Accordingly, the goal is to generate enriched code slices that precisely encapsulate statements pertinent to the patch while filtering out extraneous code.

\textbf{Patch Localization}. 
This stage commences by decompiling the target function $F_{t}$ into a pseudocode representation, $P_{t}$. The subsequent objective is to locate the corresponding code slices within $P_{t}$, leveraging the source-level slices, $C_{vul}$ and $C_{patch}$, extracted in the preceding phase. Drawing upon the demonstrated efficacy of LLMs in code comprehension tasks such as decompilation~\cite{llm4decompile} and function summarization~\cite{degpt}, we employ an LLM-based approach for this localization. Specifically, we construct a prompt that provides the LLM with the decompiled function $P_{t}$ and the source slices $C_{vul}$ and $C_{patch}$. The model is instructed to map each statement from the source slices to its corresponding sequence in the pseudocode. This process yields the final pseudocode slices, $P_{vul}$ and $P_{patch}$. Although this step identifies the pseudocode sections with the highest semantic similarity to the source, further verification is necessary to determine whether a given slice represents the vulnerable logic or the applied patch.

\textbf{Code2SMT Reasoning Engine}.
To resolve the ambiguity from the localization phase, the Code2SMT reasoning engine performs neuro-symbolic reasoning over the identified pseudocode slices, $P_{vul}$ and $P_{patch}$. It instantiates the neural-perceives, symbolic-decides principle through three stages. The neural side comprises an Extractor that deconstructs the pseudocode into its fundamental logical components and a Normalizer that translates these components into the standardized SMT-LIB format and aligns symbols across representations. The symbolic side is a Verifier that processes these SMT-LIB formulas and checks for logical equivalence between the vulnerability's logic and the code present in the target. Based solely on the solver's deterministic result, the engine produces a definitive output, classifying the binary as either Patched or Vulnerable.

\subsection{Patch Enhancement}

We adopt three main techniques to enhance patch-related code slices: data flow analysis, control flow analysis, and macro analysis.
Given a patch commit diff file and the function source code before and after the patch ($S_{vul}$ and $S_{patch}$), the module produces the corresponding patch-related slices for both versions ($C_{vul}$ and $C_{patch}$). In line with MVP \cite{MVP}, \myname employs lightweight slicing via Joern \cite{joern}, which enables a compile-free test for patch presence. This approach is restricted to intra-function analysis.

\subsubsection{Data Flow Analysis}

We first enhance patches by analyzing their data dependencies. Variables defined or used in the patch often link the patch to the rest of the program. Because patches usually change only a small part of the code, the wider roles of these variables are often overlooked. To resolve this, we perform systematic data flow analysis to recover their full behavior in the program.

\begin{figure}

\centering
    \includegraphics[width=\linewidth]{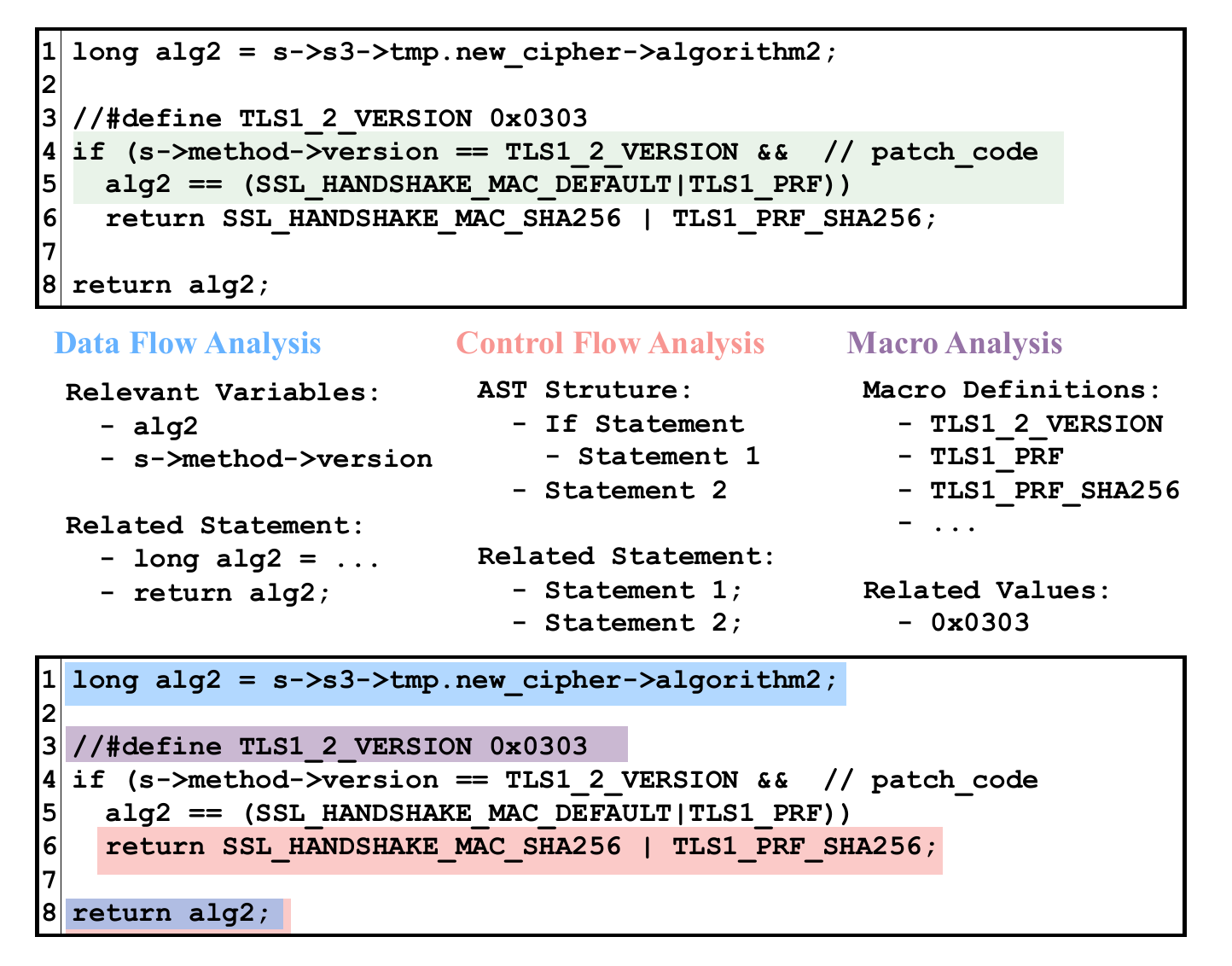}

\caption{The Patch Enhancement Module.}
\label{fig:patch_enhance}

\end{figure}

We parse the patch to collect all defined and used variables. These variables mark the main interaction points between the patch and the surrounding code. For example, as shown in Figure~\ref{fig:patch_enhance}, a change to a single \texttt{if} statement involves the condition of the \texttt{if} clause and any variables referenced in its body, which may interact with other parts of the program.

After extracting these variables, we analyze the function or module that contains the patch to locate further statements that define or use them. For instance, in Figure 4, lines 1 and 8 are data flow-related statements linked to the patched statement and provide added context for analysis.

\subsubsection{Control Flow Analysis}

While data flow analysis focuses on variable interactions, control flow analysis examines the structural and logical organization of the code affected by the patch. This analysis is essential for understanding how the patch changes execution paths, especially in the presence of loops and conditionals. By incorporating these dependencies, control flow analysis provides a clear view of the patch's effect on execution.

For each patch, we identify the enclosing control flow structure. This includes locating the entry points of control structures, such as the condition of an \texttt{if} statement or the header of a loop. We then extract the first statement of each block and the statement that follows the block. For example, if a patch changes the body of an \texttt{if} statement, we also include the corresponding \texttt{else} branch, if present, and the statement after the \texttt{if-else} block.

\subsubsection{Macro Analysis}

In addition to data flow and control flow, the third strategy addresses macro definitions, which often encode constants, expressions, or inline code. Because the target binary lacks these definitions, we resolve the concrete values of macros in the patch to align with the values present in the pseudocode.

Once the relevant macros are identified, we retrieve their values or code fragments from the source and substitute them into the patch, replacing macro names with their concrete forms. For example, in Figure~\ref{fig:patch_enhance}, the macro \texttt{TLS1\_2\_VERSION} resolves to \texttt{0x0303}, which supports correct matching in later steps (e.g., mapping 0x303 to 771 through macros). Although direct lookup may introduce errors, this lightweight method works well in practice and does not require compilation.

\subsection{Patch Localization}

The Patch Localization module aims to locate the matching pseudocode slices ($P_{vul}$ and $P_{patch}$) within the target function's pseudocode by using the enhanced patch code slices ($C_{vul}$ and $C_{patch}$) produced by the previous module. Instead of supplying only the patch slice, we provide the full function source code. To facilitate identification, each line of the patch slice in the function source is annotated with the comment ``\texttt{//patch\_code}".

\subsubsection{Truncation of Function}

To cope with the token limits of LLMs, we perform AST-guided truncation that preserves key context while keeping both syntax and semantics intact.

For the patch code, we select the first patched statement as an anchor and expand upward and downward to include nearby lines. Truncation is aligned with AST structural boundaries to maintain semantics. For instance, if the patch alters a statement inside a loop, we include the entire loop body. Expansion proceeds until a preset token budget is met. The resulting patch code is labeled ``\texttt{patch code}" for later processing. We apply a similar method to the target function's pseudocode. Instead of extracting a single statement with its context, we collect multiple slices that are truncated at AST boundaries and provide them to \myname for subsequent processing.

\subsubsection{Prompt Construction}

After truncating the patch code and pseudocode, we prepare prompts for the LLM to perform patch localization. The prompt follows a fixed template, shown in Appendix. The template contains placeholders for the patch code and one pseudocode segment, which we replace with the truncated patch code and a single pseudocode segment. The prompt asks the LLM to find the pseudocode slice that corresponds to the patch code slice.

We submit the prompt to the LLM. The model analyzes the patch code and the pseudocode segment and returns a JSON object that specifies the location of the patch slice within the pseudocode. This JSON includes the matched pseudocode slice and its location. The output yields an exact mapping between the enhanced patch code slice and the corresponding pseudocode slice, which serves as the basis for the subsequent \textit{patch verification} module.

\subsubsection{Reverse-Matching Strategy for Add and Delete Patches}

Not all vulnerabilities identified in the \textit{patch localization} module yield two matched pseudocode slices. For edit-type patches, the unique code before and after the change can each be matched to the target pseudocode, producing two distinct results for subsequent verification. In contrast, add-only and delete-only patches produce a single match between the added or removed slice and the pseudocode, making it difficult to decide whether the matched pseudocode aligns with the pre-patch or post-patch version. To resolve this, we introduce a reverse-matching strategy that guarantees two matches. The aim is to compare the pseudocode with the two source versions surrounding the patch. Because the differences introduced by the patch are small, an LLM may not reliably isolate a single pre- or post-patch slice. Therefore, for each vulnerability, \myname matches the most similar pseudocode slice to the pre-patch source and another to the post-patch source, then chooses the correct pair to achieve accurate localization.

For add-only patches, besides $P_{patch}$ that matches the added slice ($C_{patch}$), the second result $P_{vul}$ is obtained by reverse matching the pseudocode slice $P_{patch}$ against the pre-patch source $S_{vul}$. This reveals whether the pseudocode slice comes from the new code or from existing lines in the pre-patch version. Likewise, for delete-only patches, $P_{vul}$ is matched with $S_{patch}$ to derive $P_{patch}$ for verification.

\subsection{Code2SMT Reasoning Engine}

\begin{figure}

\centering
    \includegraphics[width=\linewidth]{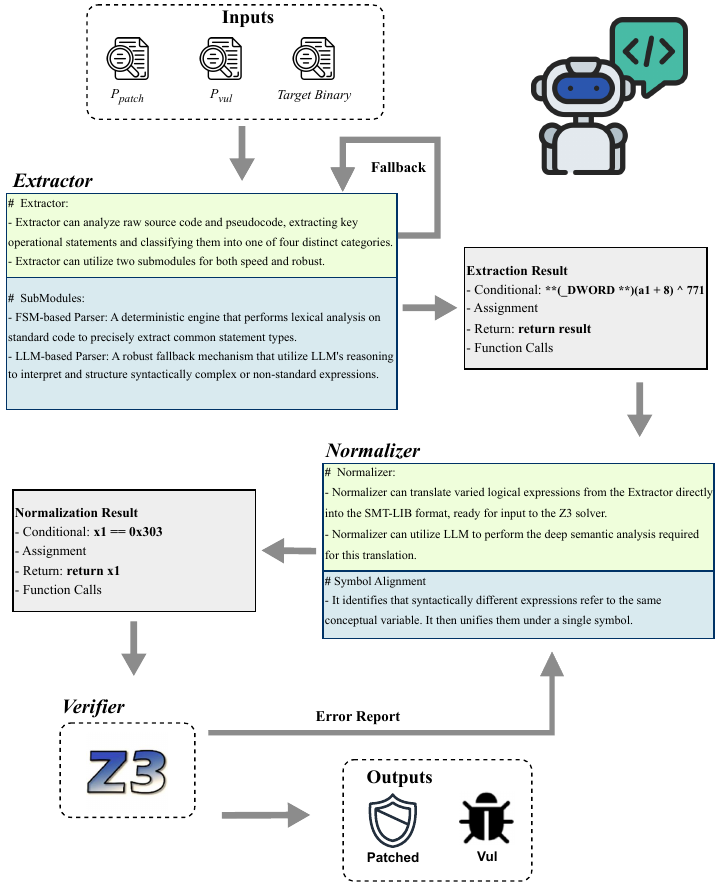}

\caption{Code2SMT Reasoning Engine.}
\label{fig:agent}
\end{figure}

The purpose of the patch verification module is to determine whether the pseudocode slice ($P_{vul}$ and $P_{patch}$) identified by the patch localization module truly corresponds to the vulnerable code or the patched code ($C_{vul}$ and $C_{patch}$). While the patch localization module identifies the pseudocode slice most similar to the patch code, additional analysis is required to distinguish whether the match aligns with the vulnerable or patched version. The Patch Verification module takes as input two locating results for the target function: one from the pre-patch version and one from the patched version. Its output is a determination of which locating result is correct.

Existing methods, including Lares, tend to invoke the Z3 solver and the LLM as two independent equivalence judges. However, since feature extraction is performed on heterogeneous source code and pseudocode, it is difficult to fully obtain semantic formulas, which frequently causes the symbolic channel to fail and forces an unreliable fallback to the neural channel. To address this, we organize the engine as a neuro-symbolic pipeline rather than two competing judges: a neural Extractor and Normalizer transform raw source code and pseudocode into a canonical, verifiable SMT-LIB format, and a symbolic Verifier renders the final decision. The LLM is deliberately confined to feature extraction and normalization so that the deterministic solver, not the LLM, makes every verdict, maximizing the use of formal methods while avoiding hallucinated conclusions.

This verification process is fully automated and forms the final step in the pipeline, enabling accurate and efficient analysis of software patches.

\subsubsection{Extractor}
The goal of Extractor, the first stage of neural perception, is to parse both well-structured source code and potentially unstructured pseudocode into a consistent intermediate representation (IR) of candidate logical expressions. It operates with a primary deterministic sub-module and an LLM-based fallback, ensuring both precision and robustness.

\textbf{FSM-based Parser Sub-module.}
In order to extract semantic formulas from code slices and perform SMT-Based constraint solving, we need to first normalize them. 
The process begins with a customized lexical analysis of the matched code slices, where the code is parsed into fundamental lexical units. Finally, we extract the normalized equation from the vocabulary unit.

We use a custom finite state machine (FSM) for lexical analysis. We take each line of code as input, provide it to the FSM character by character, and perform state transition according to each new character. This recursive definition ensures that the FSM processes the input string sequentially, one character at a time, maintaining a systematic and deterministic approach to lexical analysis.

After lexical analysis, we get the lexical units corresponding to the code. From these units, we extract four primary types of code slices:

\begin{itemize}
    \item \textbf{Conditional Statements:} e.g., \texttt{if (x > 0)}, \texttt{while (y == 1)}.
    \item \textbf{Assignment Statements:} e.g., \texttt{x = a + b}, \texttt{y = 0}.
    \item \textbf{Return Statements:} e.g., \texttt{return z}.
    \item \textbf{Function Calls:} e.g., \texttt{foo(x, y)}, \texttt{bar(x+14)}.
\end{itemize}

\textbf{LLM-based Parser Sub-module.}
For the pseudocode slice, which often defies traditional parsing, the engine utilizes this robust neural fallback. The FSM parser would fail on a complex expression like \texttt{if (**(DWORD **)(a1 + 8)}\textasciicircum{}\texttt{771)}, as the nested pointer dereferencing and type casting are non-standard. When such a failure is detected, the engine tasks an LLM to analyze the unstructured text. Leveraging its powerful reasoning capabilities, the LLM infers the underlying semantics, specifically a comparison between a value at a calculated memory address and the constant 771. Crucially, the LLM does not decide patch presence here; it only abstracts this understanding into the standardized dictionary format used by our system, ensuring consistency for subsequent processing stages.

\subsubsection{Normalizer}
Normalizer is the second stage of neural perception. It takes the heterogeneous logical expressions from the Extraction Module and unifies them into a canonical, solver-friendly format. A specialized LLM normalizer processes the IR, as this task requires a deep semantic understanding that surpasses the capabilities of rule-based systems. Its primary function is to translate the varied expressions into a standardized domain-specific language (DSL) compatible with SMT-LIB, producing the specification-compliant input that the symbolic reasoner consumes.

The most critical capability of the normalizer is Symbol Alignment, which is the linchpin that makes the neural-to-symbolic hand-off sound. Continuing the example from CVE-2013-6449, the Extraction Module produced two distinct logical fragments: one for the source code's \texttt{s->method->version} and another for the pseudocode's \texttt{**(DWORD **)(a1 + 8)}. A human expert can recognize that these refer to the same conceptual variable, but they are syntactically disparate. The LLM normalizer is tasked with inferring this semantic correspondence. It aligns both expressions to a single canonical symbol, such as \texttt{x1}. Consequently, the source code condition is normalized to \texttt{x1 == 0x303}, and the pseudocode condition becomes \texttt{x1 == 771}. This alignment is crucial, as it transforms the problem from comparing two syntactically different statements into comparing two simple equations with a shared variable, a task perfectly suited for an SMT solver. The final output is a structured, verifiable representation, ready for the symbolic decision stage.

\subsubsection{Verifier}
Verifier is the symbolic decision stage: it executes the formal equivalence check and incorporates a self-correction loop to handle solver errors, delivering the final, deterministic verdict.

The engine feeds the normalized logical representations into the Z3 SMT solver to check for logical equivalence. This process is embedded within a feedback-driven refinement loop that keeps the neural and symbolic components coupled. If Z3 returns a syntax or type error, the engine does not immediately fail. Instead, it captures the error message and uses it as feedback for the LLM-based Normalization stage. The engine prompts the LLM to correct the SMT-LIB script based on Z3's specific error, and then resubmits the corrected script for verification. This iterative self-correction lets the symbolic solver's diagnostics guide the neural normalizer, enhancing the robustness of the verification process without ever delegating the verdict to the LLM.



\section{Implementation}

\myname is implemented in approximately 3,500 lines of Python code, integrating several tools for its functionality.

For binary analysis, we use IDA Pro \cite{IDA_Pro} to decompile binaries and extract function-level pseudocode via custom scripts. IDA Pro was selected for its robust decompilation and scripting capabilities. For source code, Tree-Sitter \cite{treesitter} parses code into functions with AST-based segmentation, while Joern \cite{joern} performs advanced static analysis, including data flow and control flow, using its code property graph model.

The final, symbolic step of our pipeline involves formal verification using the Z3 SMT solver \cite{z3}, which determines the semantic equivalence of the translated code slices. The crucial neural Code2SMT translation process is powered by our reasoning engine, which leverages Claude-3.5-Sonnet. Specifically, the LLM is employed within the Extractor as a fallback parser for syntactically complex code and within the Normalizer to perform deep semantic analysis and symbol alignment, but is never used to issue the final verdict. We selected Claude-3.5-Sonnet to balance advanced reasoning capabilities with operational cost, which averages approximately \$0.30 per verification case. We use the default values for LLM parameters (temperature=1.0).

\section{Evaluation}
We evaluate the effectiveness of \myname by answering the following research questions.



\textbf{RQ1:} Does \myname improve formal verification coverage by routing neural outputs into the symbolic reasoner?

\textbf{RQ2}: How does \myname perform in the cross-optimization patch presence testing task compared to existing methods?

\textbf{RQ3}: How does \myname perform in the cross-architecture patch presence testing task? 

\textbf{RQ4}: How does every component in \myname affect the overall performance?

\textbf{RQ5}: How efficient is \myname? 

\paragraph{Experiment setup}
All experiments were conducted on a server running a 64-bit Ubuntu 22.04 LTS system. The server is equipped with an Intel(R) Xeon(R) Platinum 8460Y+ CPU, features 2.0TB of system memory, and eight NVIDIA H100 GPUs (80GB HBM3).

\paragraph{Evaluation metrics}
We adopt the evaluation criteria used in prior work~\cite{robin, ps3, react}, employing precision, recall, and F1 score to assess \myname and the baselines. Specifically, $TP$, $FP$, and $FN$ denote the numbers of patched functions correctly identified as patched, vulnerable functions incorrectly labeled as patched, and patched functions incorrectly labeled as vulnerable, respectively.

\paragraph{Dataset}
We evaluate on four well-known real-world projects from different application domains, consistent with prior work~\cite{ps3, react}. These projects cover protocol encryption, packet processing, XML analysis, and font rendering. For each collected vulnerability, we compile both vulnerable and patched versions of OSS using gcc v9.4.0 and clang v6, across optimization levels O0–O3 and architectures x86, x64, and ARM. Table~\ref{tab:dataset} reports project statistics.
In total, we collect 73 CVEs with 224 distinct vulnerable and patched functions, yielding 20 * (112 + 112) = 4480 test cases.

We also extend the dataset with 20 vulnerabilities (400 cases) disclosed after October 2024, to ensure they are absent from the training data of the model and avoid data contamination.

\begin{table}
  \renewcommand{\arraystretch}{1.2} 
  \centering
  \caption{Dataset}

  \label{tab:dataset}
   \resizebox{1.0\linewidth}{!}{
  \begin{tabular}{cccccc}
    \toprule
     Project & CVE & $func_v$ & Version & Binary & Testcase\\
    \midrule
    OpenSSL & 34 &  100 & 20 & 400 & 2000\\
    Freetype & 7 &  18 &  5 & 100 & 360\\
    Tcpdump & 25 &  78 &  2 & 40 & 1560\\
    Libxml2 & 7 &  28 &  4 & 80 & 560\\
    \midrule
    All & 73 &  224 & 31 & 620 & 4480\\
  \bottomrule
\end{tabular}}

\end{table}

\paragraph{Baselines}

We select three baselines in our experiments.

\begin{itemize}
    \item \textbf{BinXray \cite{BinXray}}: A classic syntactic-based approach for patch presence testing, widely adopted in various works.
    \item \textbf{Robin \cite{robin}}: A state-of-the-art semantic-based patch presence testing approach that leverages symbolic execution to extract patch features.
    \item \textbf{PS3 \cite{ps3}}: A state-of-the-art patch presence testing approach specifically designed to address challenges posed by compiler optimization.
\end{itemize}

Additionally, REACT \cite{react} needs to be compiled with clang to generate Intermediate Representations (IR). This process requires an additional compilation process after the normal compilation, which is more difficult to use than other methods. Since we aim to implement a compile-free method, REACT is orthogonal to our scope of consideration.

\subsection{Formal Verification Coverage (RQ1)}

In Lares, the symbolic channel (Z3) frequently failed when presented with complex code patterns, unhandled syntactic structures, or misaligned variables, severely limiting its applicability and forcing a fallback to the less reliable neural channel for the final decision. We evaluate the improvement \myname achieves in successfully translating different types of code into the SMT-LIB format, i.e., in handing every case to the symbolic reasoner.

We compared the formal verification coverage of \myname against the original Lares on our entire dataset of 4480 samples. We define ``coverage" as the percentage of test cases that the Z3 solver could successfully and completely process to reach a definitive conclusion.

The results demonstrate a dramatic improvement in scalability. Lares could only successfully formalize 897 of the 4480 samples, achieving a verification coverage of only 20\%. Its simplistic translator failed on complex control flows and data structures, so the remaining 80\% degenerated to neural-channel guesses. In contrast, \myname, driven by the Code2SMT reasoning engine, successfully processed all 4480 samples, achieving 100\% coverage. This is a direct result of the neuro-symbolic pipeline, which uses deterministic parsers for common patterns and LLM-driven normalization to handle complex or non-standard code, ensuring every sample is converted into a verifiable format for the symbolic reasoner.

\begin{tcolorbox}[colback=gray!10,
                  colframe=black,
                  arc=1mm, auto outer arc,
                  boxrule=1.5pt,
                 ]

\textbf{Answering RQ1:} By using the LLM as a feature extractor that feeds the symbolic reasoner, \myname improves the formal verification coverage to 100\%, representing an 80\% increase over Lares.
\end{tcolorbox}

\begin{table}[t]
\renewcommand{\arraystretch}{1.2}
\centering
\caption{Comparison of Samples Processed by Lares and SyLares}
\label{tab:lares_comparison_swapped_complex}
\resizebox{1.0\linewidth}{!}{
\begin{tabular}{cccc}
\toprule
Project & Total Testcases & Lares & SyLares \\
\midrule
OpenSSL & 2000 & 243 (12.15\%) & 2000 \\
Freetype & 360 & 229 (63.61\%) & 360 \\
Tcpdump & 1560 & 227 (14.55\%) & 1560 \\
Libxml2 & 560 & 198 (35.36\%) & 560 \\
\midrule
All & 4480 & 897 (20.02\%) & 4480 (\textbf{100\%}) \\
\bottomrule
\end{tabular}}

\small
\end{table}
\subsection{Cross-optimization Task (RQ2)}

\begin{table*}[t]
\renewcommand{\arraystretch}{1} 
\centering
\caption{Performance of patch presence testing with different optimization options on x86.}
\vspace{0.5em}
\label{table1}
\begin{tabular}{c|ccc|ccc|ccc|ccc|ccc|ccc}
\toprule[2pt]
\multicolumn{1}{c|}{\multirow{2}{*}{Model}} & \multicolumn{3}{c|}{O0}&\multicolumn{3}{c|}{O1}&\multicolumn{3}{c|}{O2}&\multicolumn{3}{c|}{O3}&\multicolumn{3}{c|}{Average}&\multicolumn{3}{c}{$Average_{new\_vul}$}\\
\cline{2-19}

\multicolumn{1}{c|}{} & P & R & F1 & P & R & F1 & P & R & F1 & P & R & F1 & P & R & F1 & P & R & F1\\
\hline


\multirow{1}{*}{BinXray} &1.0	&0.49	&0.66	&-	&-	&- &-	&-	&- &-	&-	&-	&-	&- &- &-	&- &-\\

\multirow{1}{*}{Robin} &0.64	&0.75	&0.69	&0.63	&0.70	&0.67 &0.62	&0.69	&0.66 &0.63	&0.73	&0.68 &0.63	&0.72	&0.67 & 0.69 & 0.73 & 0.70\\

\multirow{1}{*}{PS3} &0.76	&0.93	&0.83	&0.56	&0.70	&0.62 &0.56	&0.74	&0.64 &0.56	&0.74 &0.64 &0.61	 &0.78	&0.68 & 0.62 & 0.79 & 0.69\\

\multirow{1}{*}{Lares} &0.69	&0.91	&0.79	&0.71	&0.77	&0.74 &0.72	&0.79	&0.75 &0.78	&0.83	&0.81 & 0.72 &0.83	&0.77 & 0.76 & 0.81 & 0.79\\

\multirow{1}{*}{\myname} &0.79	&0.89	&0.84 	&0.86	&0.84	&0.85   &0.83	&0.81	&0.82 &0.80	&0.84	&0.82 &\textbf{0.82}	&\textbf{0.84}	&\textbf{0.83} & 0.81 & 0.83 & 0.82\\

\bottomrule[2pt]
\end{tabular}
\label{table_MAP}
\end{table*}

\begin{table*}[t]
\renewcommand{\arraystretch}{1} 
\centering
\caption{Patch presence testing accuracy of different compilation environment.}
\vspace{0.5em}
\label{table2}
\begin{tabular}{c|ccc|ccc|ccc|ccc|ccc|ccc}
\toprule[2pt]
\multicolumn{1}{c|}{\multirow{2}{*}{Enviroment}} & \multicolumn{3}{c|}{O0} & \multicolumn{3}{c|}{O1} & \multicolumn{3}{c|}{O2} & \multicolumn{3}{c|}{O3} & \multicolumn{3}{c|}{Average} & \multicolumn{3}{c}{$Average_{new\_vul}$} \\
\cline{2-19}
\multicolumn{1}{c|}{} & P & R & F1 & P & R & F1 & P & R & F1 & P & R & F1 & P & R & F1 & P & R & F1 \\
\hline

\multirow{1}{*}{x86-gcc} 
& 0.70 & 0.87 & 0.78
& 0.75 & 0.81 & 0.78
& 0.72 & 0.79 & 0.75
& 0.77 & 0.82 & 0.79
& 0.74 & 0.82 & 0.78
& 0.79 & 0.83 & 0.81\\

\multirow{1}{*}{x64-gcc} 
& 0.69 & 0.87 & 0.77
& 0.73 & 0.81 & 0.77
& 0.70 & 0.81 & 0.75
& 0.77 & 0.83 & 0.80
& 0.72 & 0.83 & 0.77
& 0.79 & 0.84 & 0.81\\

\multirow{1}{*}{arm-gcc} 
& 0.66 & 0.82 & 0.73
& 0.68 & 0.77 & 0.72
& 0.72 & 0.77 & 0.75
& 0.74 & 0.78 & 0.76
& 0.70 & 0.79 & 0.74
& 0.72 & 0.76 & 0.74\\

\multirow{1}{*}{x86-clang} 
&0.79	&0.89	&0.84 	&0.86	&0.84	&0.85   &0.83	&0.81	&0.82 &0.80	&0.84	&0.82  &0.82 &0.84	&0.83 & 0.81 & 0.83 & 0.82\\

\multirow{1}{*}{x64-clang} 
& 0.71 & 0.92 & 0.80
& 0.72 & 0.79 & 0.75
& 0.73 & 0.78 & 0.75
& 0.76 & 0.82 & 0.79
& 0.73 & 0.83 & 0.78
& 0.77 & 0.83 & 0.80\\

\bottomrule[2pt]
\end{tabular}
\label{table_MAP2}
\end{table*}

To evaluate \myname' performance in the cross-optimization (O0, O1, O2, and O3) patch presence testing task, we compare it against three state-of-the-art methods. The results are summarized in Table~\ref{table_MAP}.

BinXray~\cite{BinXray} achieves perfect precision at O0 (1.0) but fails at higher optimization levels (O1, O2, O3) due to its reliance on syntactic features. Robin~\cite{robin} is relatively stable across optimization levels, with F1 scores between 0.66 and 0.69, but its precision and recall are lower than those of Lares. The main reasons are failures in symbolic execution for some functions and feature sets that do not sufficiently distinguish patched from unpatched code. PS3~\cite{ps3} performs well at O0 (F1 = 0.83) but degrades under higher optimizations (F1 = 0.62, 0.64, and 0.64 at O1, O2, and O3), mainly because many features it uses are unrelated to the patch, adding noise and causing missed detections. Lares demonstrates robust performance across all optimization levels, with F1 scores of 0.79 (O0), 0.74 (O1), 0.75 (O2), and 0.81 (O3). Its average F1 score of 0.77 is a notable improvement over other baselines, particularly PS3 at higher levels of optimization. 

\myname outperforms all baseline methods, demonstrating highly consistent and superior performance across all optimization levels with an average F1 score of 0.83. Unlike BinXray~\cite{BinXray}, which fails at O1-O3, \myname effectively handles optimized code, maintaining F1 scores at 0.82 or higher across all levels. Compared to PS3~\cite{ps3}, which degrades significantly under optimization (e.g., F1=0.62 at O1), \myname shows remarkable stability, with its F1 score peaking at 0.85 (O1). \myname also significantly surpasses Robin~\cite{robin} (Avg F1=0.67) and Lares (Avg F1=0.77). It achieves a higher F1 score than Lares at every optimization level and maintains higher average precision (0.82 vs. 0.72) and recall (0.84 vs. 0.83), confirming its robust patch detection capabilities.

Specifically, we evaluated 400 new cases, denoted as $Average_{new\_vul}$. The results on these vulnerabilities align with those from the conventional datasets, and \myname shows a substantial improvement over existing methods. This further indicates that \myname's gains are not caused by data contamination. This matches our intuition: \myname relies on the LLM’s code analysis ability rather than memorized knowledge of vulnerabilities in the training set, and thus is not affected by data contamination.

\begin{tcolorbox}[colback=gray!10,
                  colframe=black,
                  arc=1mm, auto outer arc,
                  boxrule=1.5pt,
                 ]

\textbf{Answering RQ2:} \myname can handle diverse optimization levels and significantly improve the precision and recall of existing methods.
\end{tcolorbox}

\subsection{Cross-architecture Task (RQ3)}

To assess \myname in cross-architecture and cross-compiler patch presence testing, we evaluate it on three architectures (x86, x64, ARM) and two compilers (GCC and Clang) across multiple optimization levels. Table~\ref{table_MAP2} reports precision, recall, and F1 under these settings.

Existing approaches such as BinXray \cite{BinXray}, PS3 \cite{ps3}, and Robin \cite{robin} depend on architecture- or compiler-specific features, which limits their applicability in cross-architecture or cross-compiler scenarios. Their features are tightly tied to instruction tokens, and instruction sets differ by architecture. In particular, Robin and BinXray can extract features for ARM, but they require the target and reference binaries to share the same architecture and do not support cross-architecture detection. Consequently, they are confined to single-architecture or single-compiler settings, precluding direct comparisons. We therefore report only \myname. In contrast, \myname is designed to generalize across both architectures and compilers.

\myname, which corresponds to the high-performing x86-clang configuration, sets the benchmark with an average F1 score of 0.83. It demonstrates remarkable stability across all optimization levels, posting F1 scores of 0.84 (O0), 0.85 (O1), 0.82 (O2), and 0.82 (O3). This performance not only represents the peak in this comparison but also significantly surpasses the GCC-based configurations (0.74-0.78 F1) and the x64-clang setup (0.78 F1). With high average precision (0.82) and recall (0.84), \myname demonstrates superior, reliable patch detection capabilities in this setting.

\begin{tcolorbox}[colback=gray!10,
                  colframe=black,
                  arc=1mm, auto outer arc,
                  boxrule=1.5pt,
                 ]

\textbf{Answering RQ3:} \myname outperforms existing methods by effectively handling the cross-architecture and cross-compiler patch presence testing task.
\end{tcolorbox}

\subsection{Ablation Evaluation (RQ4)}

We evaluate \myname using different LLMs and temperature settings. For nonzero temperatures, each case is repeated five times, and the results show no notable variation. As reported in Table~\ref{tab:llm}, Claude-3.5-Sonet attains the highest F1 score, with GPT-4o next. The open-source DeepSeek-Coder achieves the highest Recall but the lowest Precision. Claude-3.5 and GPT-4o markedly surpass existing methods, while other models deliver comparable results. More advanced models, such as o1, are excluded due to cost constraints.

To evaluate the contribution of each module in \myname, we conducted an ablation study by progressively removing individual components. The results are summarized in Table~\ref{tab:freq4}, where the precision, recall, and F1 score of different variants of \myname are reported. The settings are same as Table~\ref{table_MAP}.

\begin{table}[t]
\centering
\renewcommand{\arraystretch}{1.2}
\caption{Performance of different LLMs across settings. Darker green indicates better performance}
\label{tab:llm}
\resizebox{1.0\linewidth}{!}{
\begin{tabular}{
    >{\raggedright\arraybackslash}p{0.1cm}
    >{\raggedright\arraybackslash}p{2.5cm}!{\hspace{0.3pt}\vspace{0.3pt}}
    *{4}{>{\centering\arraybackslash}p{0.8cm}!{\hspace{0.3pt}} }
}
\multicolumn{2}{c}{} &
\multicolumn{4}{c}{\textbf{Temperature}} \\
\addlinespace[3pt]
\multicolumn{2}{c}{\textbf{Metrics, Model} } &
\cellcolor{lightgray}\textbf{0.0} &
\cellcolor{lightgray}\textbf{0.5} &
\cellcolor{lightgray}\textbf{0.7} &
\cellcolor{lightgray}\textbf{1.0} \\
\addlinespace[3pt]

\multirow{5}{*}{\rotatebox[origin=c]{90}{\textbf{Precision}}}
  & \cellcolor{lightgray} {\small GPT-3.5-Turbo}
    & 0.553 & 0.564 & 0.586 & 0.566 \\
  & \cellcolor{lightgray} {\small DeepSeek-Coder}
    & 0.547 & 0.552 & 0.558 & 0.579 \\
  & \cellcolor{lightgray} {\small GPT-4o}
    & 0.627 & 0.642 & \cellcolor{lightgreen!50} 0.668 & \cellcolor{lightgreen!50} 0.708 \\[2pt]
  & \cellcolor{lightgray} {\small Claude-3.5-Sonnet}
    & \cellcolor{lightgreen!50} 0.673
    & \cellcolor{lightgreen}    0.725
    & \cellcolor{green}         0.763
    & \cellcolor{green}         0.820 \\
\addlinespace[3pt]

\multirow{5}{*}{\rotatebox[origin=c]{90}{\textbf{Recall}}}
  & \cellcolor{lightgray} {\small GPT-3.5-Turbo}
    & \cellcolor{lightgreen} 0.790
    & \cellcolor{lightgreen} 0.809
    & \cellcolor{lightgreen} 0.796
    & \cellcolor{lightgreen!50} 0.768 \\
  & \cellcolor{lightgray} {\small DeepSeek-Coder}
    & \cellcolor{green} 0.963
    & \cellcolor{green} 0.906
    & \cellcolor{lightgreen} 0.913
    & \cellcolor{lightgreen} 0.920 \\
  & \cellcolor{lightgray} {\small GPT-4o}
    & \cellcolor{lightgreen} 0.881
    & \cellcolor{lightgreen} 0.919
    & \cellcolor{green} 0.947
    & \cellcolor{green} 0.945 \\[2pt]
  & \cellcolor{lightgray} {\small Claude-3.5-Sonnet}
    & 0.727
    & \cellcolor{lightgreen} 0.850
    & \cellcolor{lightgreen!50} 0.838
    & \cellcolor{lightgreen} 0.840 \\
\addlinespace[3pt]

\multirow{5}{*}{\rotatebox[origin=c]{90}{\textbf{F1 Score}}}
  & \cellcolor{lightgray}{\small GPT-3.5-Turbo}
    & 0.656 & 0.671 & \cellcolor{lightgreen!50} 0.683 & 0.655 \\
  & \cellcolor{lightgray}{\small DeepSeek-Coder}
    & \cellcolor{lightgreen!50} 0.693 & \cellcolor{lightgreen!50} 0.707 & \cellcolor{lightgreen!50} 0.704 & \cellcolor{lightgreen!50} 0.721 \\
  & \cellcolor{lightgray}{\small GPT-4o}
    & \cellcolor{lightgreen!50} 0.732 & \cellcolor{lightgreen} 0.780 & \cellcolor{lightgreen} 0.783 & \cellcolor{lightgreen} 0.809 \\[2pt]
  & \cellcolor{lightgray}{\small Claude-3.5-Sonnet}
    & \cellcolor{lightgreen!50} 0.699
    & \cellcolor{lightgreen}    0.783
    & \cellcolor{lightgreen}    0.799
    & \cellcolor{green}         0.830 \\
\addlinespace[3pt]
\end{tabular}}
\end{table}

\begin{table}[t]
\renewcommand{\arraystretch}{1} 
  \centering
  \caption{Ablation study}
  \label{tab:freq4}
  \setlength{\tabcolsep}{10pt}
  \resizebox{1.0\linewidth}{!}{
  \begin{tabular}{cccc}
    \toprule
     Model & Precision & Recall &  F1\\
    \midrule
    $SyLares_{-pe}$ & 0.697 & 0.778 & 0.736\\
    $SyLares_{-pl}$ & 0.527 & 0.642 & 0.580\\
    $SyLares_{-z3}$ & 0.624 & 0.735 & 0.676\\
    $SyLares_{-ex}$ & 0.780 & 0.800 & 0.790\\
    $SyLares_{-no}$ & 0.550 & 0.600 & 0.574\\
    $SyLares$       & 0.821 & 0.836 & 0.829\\
    \bottomrule
  \end{tabular}}
\end{table}

Removing the patch enhancement module ($SyLares_{-pe}$) reduces the system to a generic LLM-based function similarity matcher, yielding precision 0.697, recall 0.778, and F1 0.736. Because without the enriched semantics from this module, the LLM cannot reliably capture subtle patch differences.

Eliminating the patch localization module ($SyLares_{-pl}$) and asking the LLM to detect patches directly from pseudocode leads to a larger decline, with precision 0.527, recall 0.642, and F1 0.580. Without localization, the LLM must analyze raw pseudocode that often lacks variable names and contains structural ambiguities, hindering accurate decisions on patch presence. These results underscore the role of patch localization in restricting analysis to relevant slices, allowing the LLM to focus on smaller, semantically meaningful regions.

Removing the LLM-based fallback sub-module from the Extractor ($SyLares_{-ex}$) results in a performance drop, with an F1 score of 0.790 (Precision 0.780, Recall 0.800). This configuration relies solely on the deterministic FSM-based parser. The decline from the complete model's 0.829 F1 highlights the necessity of the neural fallback for perception. While the FSM parser handles well-structured code, it fails on complex or unstructured pseudocode. Without the LLM's ability to parse these non-standard cases, the system's accuracy is affected, but the F1 score remains relatively high, indicating that the FSM parser is effective for the majority of cases.

Eliminating the Normalizer module ($SyLares_{-no}$) causes a severe degradation in performance, with the F1 score plummeting to 0.574 (Precision 0.550, Recall 0.600). This sharp drop underscores the critical role of Symbol Alignment as the neural-to-symbolic bridge. Without the LLM-based normalizer, the system cannot semantically align disparate representations from source code and pseudocode. This failure to unify symbols prevents the symbolic reasoner from correctly comparing the logic, leading to a high rate of missed or incorrect patch detections.

Removing the Z3 solver and replacing it with an LLM for the final verification step ($SyLares_{-z3}$), which collapses the neuro-symbolic design back into a purely neural decision, results in a significant performance drop, with an F1 score of 0.676 (Precision 0.624, Recall 0.735). This substantial decline from \myname's 0.829 F1 score demonstrates that delegating the final verdict to the symbolic solver is crucial for achieving high confidence and accuracy. Without its deterministic verification of complex logical equivalence, the system's overall performance is clearly affected, confirming that letting the symbolic component decide is a critical design choice rather than merely complementary to the LLM's reasoning.

\begin{figure}
\centering

    \includegraphics[width=1.0\linewidth]{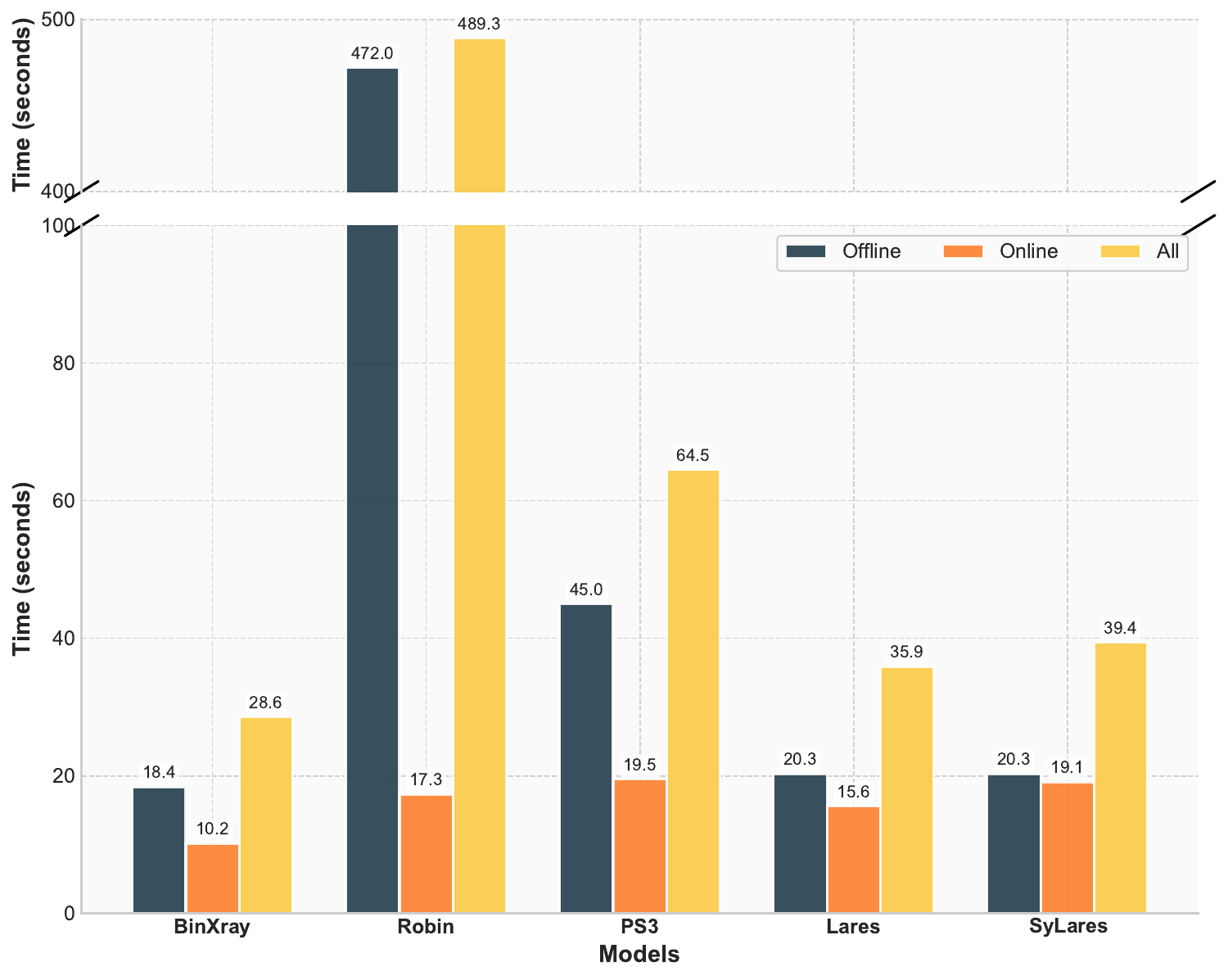}

\caption{Timecost Evaluation (seconds).}
\label{fig:timecost}
\end{figure}

\begin{tcolorbox}[colback=gray!10,
                  colframe=black,
                  arc=1mm, auto outer arc,
                  boxrule=1.5pt,
                 ]

\textbf{Answering RQ5:} Leveraging LLM, \myname significantly surpasses existing methods. Each component contributes critically to its overall performance.
\end{tcolorbox}

\subsection{Efficiency (RQ5)}

To assess \myname’s efficiency, we compare its runtime with existing methods. The runtime is split into two phases: \emph{OffLine}, which covers preprocessing such as reverse engineering and patch feature extraction, \emph{OnLine}, which is the per-binary patch presence testing time. Figure~\ref{fig:timecost} shows the results.

BinXray \cite{BinXray} is the fastest, with a total runtime of 28.6 seconds, because it relies on simple syntactic features. Robin \cite{robin} is the slowest at 489.3 seconds, with the \emph{OffLine} phase (472.0 seconds) accounting for most of the time due to symbolic execution for feature extraction. PS3 \cite{ps3} is more efficient than Robin by replacing symbolic execution with symbolic simulation, reducing the \emph{OffLine} time to 45.0 seconds for a total time of 64.5 seconds. Lares is also highly efficient, with a total runtime of 35.9 seconds. It avoids heavy computational overhead, resulting in a fast \emph{OffLine} phase (20.3 seconds) and an \emph{OnLine} phase (15.6 seconds).

\myname shows a similarly efficient profile, with a total runtime of 39.4 seconds. This is dramatically faster than the heavyweight approaches of Robin \cite{robin} (489.3s) and PS3 \cite{ps3} (64.5s). It shares the same \emph{OffLine} phase time with Lares at 20.3 seconds, demonstrating the same feature-extraction efficiency. Its \emph{OnLine} phase takes 19.1 seconds. This is slightly longer than Lares’ 15.6 seconds because \myname introduces a Code2SMT reasoning engine; both its Extractor and Normalizer components require multiple calls to the LLM during this phase. This modest increase in online processing has minimal impact on the total runtime, which remains highly efficient and significantly outperforms heavyweight methods.

The runtime comparison in Figure~\ref{fig:timecost} does not include the time required for compiling binaries. BinXray \cite{BinXray}, Robin \cite{robin} and PS3 \cite{ps3} require a compilation process to construct patch features, which adds significant overhead in real-world use.

\begin{tcolorbox}[colback=gray!10,
                  colframe=black,
                  arc=1mm, auto outer arc,
                  boxrule=1.5pt,
                 ]

\textbf{Answering RQ5:} \myname outperforms existing methods and particularly suitable for scenarios where speed and adaptability are crucial.
\end{tcolorbox}

\subsection{Failure Cases Analysis}
We investigate various failure cases in this experiment and identify several potential causes behind these inaccuracies:

\begin{itemize}
    \item \textbf{Repeated patch code}: Patch-introduced statements may already exist elsewhere in the vulnerable function, which can mislead the classifier. For example, in CVE-2014-3510, the \texttt{if} statement ``if ($\mathtt{s \rightarrow session \rightarrow sess_{cert} == NULL}$)" and its body also appear in another location of the vulnerable function. As this is a common assertion, \myname also detects it there and incorrectly labels the vulnerable function as patched. This case is uncommon in our dataset, so we did not add special handling to preserve generality. Future work can enhance the Patch Enhancement module by incorporating more context to mitigate this failure mode.
    \item \textbf{Stealthy patches}: Stealthy patches make very small edits. Differences across compiler environments can mask these changes, complicating accurate decisions. These cases are challenging even for human analysts. Concretely, the LLM attains high precision on add/delete-type patches but can fail on edit-type patches by overlooking subtle edits or confusing them. \myname improves robustness on edit-type patches and uniquely resolves cases in long functions, where the LLM alone tends to mislocalize.
    
    \item \textbf{The Hallucinations of LLM}: While \myname significantly mitigates issues from direct LLM reasoning by confining the LLM to feature extraction, some risk of hallucination remains. This risk is now concentrated in the Normalizer step of the Code2SMT reasoning engine rather than in the final verdict. Hallucinations at this stage can cause the LLM to generate an SMT-LIB translation that is not semantically equivalent to the input pseudocode, which could lead to incorrect verification results by the solver. Future work will mitigate this risk by fine-tuning a specialized LLM to enhance its capability for this task.
\end{itemize}

We engaged three code analysis experts to examine the failure cases. They found that by routing neural perception into the symbolic reasoner for the final decision, \myname solves the majority of hallucination problems caused by the neural-channel reasoning used in Lares. This reduced the proportion of judgment errors caused by hallucinations from 50\% to approximately 20\% of failures. Overall, \myname shows that neuro-symbolic reasoning has strong potential for patch presence testing. Although hallucinations prevent perfect decisions in certain cases, these issues can be mitigated by fine-tuning a specialized LLM to improve task-specific capability.
\section{Discussion}


Lares, as the first compilation-free approach in this area, showed promise but was ultimately unreliable. We also investigated CLAP \cite{clap}, the most related LLM-based method for binary code similarity. However, it completely failed to distinguish between vulnerable and patched functions, classifying nearly all functions as vulnerable. In contrast, \myname provides truly interpretable outputs. Where Lares often failed in its symbolic channel and fell back to unreliable neural-channel reasoning, \myname's neuro-symbolic Code2SMT reasoning engine successfully formalizes all cases into Z3 equations, providing a formal proof of patch presence that serves as a valuable, verifiable reference.

In this work, we employed a zero-shot commercial LLM (Claude-3.5) as the neural component of our reasoning engine and achieved promising results. This approach, however, incurs an API usage cost. A promising avenue for future work is fine-tuning open-source LLMs on patch-related data, which could further enhance detection performance while reducing deployment costs. Occasionally, the LLM components within the engine may return results in an unexpected format due to LLM randomness, requiring a retry. This happens rarely and does not significantly impact the robustness of \myname.

Our dataset includes a diverse set of vulnerable and patched functions from various projects. This dataset is consistent with previous work \cite{ps3} but goes beyond it by generating significantly more test cases, totaling 4480 samples. This larger dataset ensures a more thorough and reliable evaluation of \myname' performance, covering a wide range of real-world scenarios and challenges.
\section{Related Work}

\subsection{Vulnerability Detection}

Binary code similarity detection is fundamental \cite{Code_Similarity_survey1, Code_Similarity_survey2} to tasks such as malware analysis \cite{BinSim}, vulnerability detection \cite{QF_bugsearch, IR-Cross_arch_bug_search}, and software reuse identification \cite{ISRD, LibDB, LibDI, LibDX}. Our work targets the detection of 1-day vulnerabilities arising from binary code reuse \cite{Vulhawk, jTrans, OSCAR, Asteria-pro}. The central challenge is to bridge differences introduced by diverse compilation environments and to recover semantic equivalence.

Recent work applies Natural Language Processing (NLP) and structural analysis to binary similarity detection. NLP-based methods, including SAFE \cite{SAFE}, Trex \cite{Trex}, and CEBin \cite{cebin}, model code as natural language and learn semantic embeddings for matching. Structural methods such as Gemini \cite{Gemini, libvdiff}, ASTeria \cite{Asteria}, and HermesSim \cite{HermesSim} leverage control flow graphs (CFG), abstract syntax trees (AST), or tailored semantic graphs to detect similarity.

LLMs have recently become a new paradigm in this area. Approaches such as CLAP \cite{clap} and SCALE \cite{scale} employ LLMs to improve similarity detection. However, these approaches use LLMs as end-to-end discriminators, inheriting their non-determinism. \myname instead follows a neuro-symbolic reasoning paradigm that couples the LLM's generalized perception with the deterministic reasoning of a symbolic solver, using the LLM only to produce specification-compliant inputs and leaving the final verdict to formal methods.

Despite these advances, existing techniques have difficulty determining whether identified vulnerabilities have been patched. We address this gap by focusing on patch presence testing, which complements vulnerability detection by verifying whether a vulnerability has been mitigated, yielding a more complete assessment of software security.

\subsection{Vulnerability Verification}  
To reduce false positives, verification techniques are used, including dynamic and static methods.

Dynamic approaches, such as Directed Grey-box Fuzzing (DGF) \cite{dafl, aflplus, selectfuzz, beacon}, generate proofs of concept (PoC) by fuzzing target binaries \cite{octopocs}. AFLGo \cite{aflgo}, a representative DGF method, mutates seeds to favor paths closest to the target vulnerability. Follow-up systems, Hawkeye \cite{hawkeye}, WindRanger \cite{WindRanger}, and TransferFuzz \cite{TransferFuzz}, extend AFLGo to trigger vulnerabilities directly. But these techniques are time-consuming and apply to a limited subset of memory vulnerabilities.

Static techniques such as patch existence testing have been proposed to address these limitations. These methods analyze patch modifications to determine whether a function remains vulnerable or has been patched. Prior studies \cite{patch_survey1, patch_survey2} have examined the evolution of vulnerability patches and categorized different types of security patches. Patch existence testing for Java, Android, and C/C++ source code is relatively mature \cite{AndroidPatch}. In contrast, applying patch existence testing to C/C++ binaries poses distinct challenges due to variations in compilation environments.
BinXray \cite{BinXray} employs lightweight syntactic features. Robin \cite{robin} and PS3 \cite{ps3} use symbolic execution and symbolic simulation to improve accuracy but face scalability constraints. Kernel-oriented solutions such as Fiber \cite{Fiber} and PDiff \cite{Pdiff} have also been introduced.

Many existing methods trade accuracy for usability. We present \myname, a lightweight neuro-symbolic approach that attains both high usability and strong accuracy, while supporting cross-optimization, cross-architecture, and cross-compiler settings.

\section{Conclusion}
Determining whether a patch is present in a target binary is a crucial task for mitigating 1-day vulnerabilities, yet traditional compilation-dependent methods lack scalability and accuracy. In this paper, we presented \myname, an advanced framework that overcomes the critical limitations of our previous compilation-free approach, Lares. While Lares pioneered the joint use of LLMs and SMT solvers, it ran them as two decoupled channels, so it suffered from LLM hallucinations whenever the symbolic channel failed and the neural channel had to issue the final verdict. \myname recasts the task as neuro-symbolic reasoning: a Code2SMT reasoning engine uses the LLM purely as a neural feature extractor that produces specification-compliant SMT-LIB representations through FSM-based parsing, LLM-driven normalization, and symbol alignment, and leaves the final deterministic decision to the symbolic Z3 solver. Our evaluation demonstrates that \myname significantly enhances reliability, improving accuracy by 10\% over Lares and increasing formal verification coverage by 80\%. Compared to other state-of-the-art methods, it achieves 20\% improvement in precision and 15\% in F1 score, demonstrating its effectiveness as a robust, scalable, and automated solution for compilation-free patch presence testing.
\bibliographystyle{unsrt}
\bibliography{ref}

\end{document}